\documentclass[prb,showpacs,twocolumn]{revtex4}
\usepackage{graphics,graphicx}
\begin{document}
\title{Doping Dependence of Spin Dynamics in Electron-Doped Ba(Fe$_{1-x}$Co$_x$)$_2$As$_2$}
\author{K.~Matan$^{1,2}$}
 \email{kmatan@issp.u-tokyo.ac.jp}
 \altaffiliation[Also at ]{Department of Physics, Faculty of Science, Mahidol University, 272 Rama VI Rd., Ratchathewi, Bangkok 10400 Thailand.}
\author{S.~Ibuka$^{1,2}$}
\author{R.~Morinaga$^{1,2}$}
\author{Songxue~Chi$^{3,4}$}
\author{J.~W.~Lynn$^{3}$}
\author{A.~D.~Christianson$^{5}$}
\author{M.~D.~Lumsden$^{5}$}
\author{T.~J.~Sato$^{1,2}$}
 \email{taku@issp.u-tokyo.ac.jp}
 \affiliation{$^1$Neutron Science Laboratory, Institute for Solid State Physics, University of Tokyo, 106-1 Shirakata, Tokai, Ibaraki 319-1106, Japan\\
 $^2$TRIP, JST, 5, Sanbancho, Chiyoda, Tokyo 102-0075, Japan\\
 $^3$NIST Center for Neutron Research, National Institute of Standards and Technology, Gaithersburg, Maryland 20899, USA\\
$^4$Department of Materials Science and Engineering, University of Maryland, College Park, Maryland 20742, USA\\
 $^5$Oak Ridge National Laboratory, Oak Ridge, Tennessee 37831, USA}
\date{\today}

\begin{abstract}
The spin dynamics in single crystal, electron-doped Ba(Fe$_{1-x}$Co$_x$)$_2$As$_2$ has been investigated by inelastic neutron scattering over the full range from undoped to the overdoped regime. We observe damped magnetic fluctuations in the normal state of the optimally doped compound ($x=0.06$) that share a remarkable similarity with those in the paramagnetic state of the parent compound ($x=0$). In the overdoped superconducting compound ($x=0.14$), magnetic excitations show a gap-like behavior, possibly related to a topological change in the hole Fermi surface (Lifshitz transition), while the imaginary part of the spin susceptibility $\chi''$ prominently resembles that of the overdoped cuprates. For the heavily overdoped, non-superconducting compound ($x=0.24$) the magnetic scattering disappears, which could be attributed to the absence of a hole Fermi-surface pocket observed by photoemission.
\end{abstract}

\maketitle

\section{Introduction}
One major difference between conventional and high-T$_c$-cuprate superconductors is the proximity to a competing magnetically ordered state in the latter, and it has long been believed that magnetic fluctuations could replace the role of phonons in mediating an electron-pairing interaction.  This mechanism could give rise to more tightly bound Cooper pairs, elevating the transition temperature.  The recent discovery of iron pnictide superconductors\cite{kamihara08} with $T_c$ exceeding 50 K (Ref.~\onlinecite{RenZhiAn:2008p8209}) in close proximity to antiferromagnetic order reinvigorates this idea.

For the parent compounds of the cuprates, the magnetic properties are well described by the two-dimensional (2D) quantum non-linear sigma model,\cite{chakravarty,hasenfratz} and magnetic order is driven by a large instantaneous 2D correlation length, weak interlayer coupling, and spin anisotropies.\cite{kastner}  For the iron pnictides, on the other hand, there is still much debate over the nature of the magnetism.\cite{Diallo:2009p4224,Zhao:2009p7537}  For example, it remains controversial whether the stripe-type antiferromagnetic order in the parent compounds is stabilized by the spin-density-wave instability due to Fermi-surface nesting or by anisotropic in-plane exchange interactions due to $3d$ orbital ordering.\cite{Yildirim:2008p1163,Xu:2008p1118,Krueger:2009p7994,Phillips:2009p8891}  More importantly, spin fluctuations in the doped compounds, which are arguably a key to understand the pairing mechanism, are largely unexplored. In particular, the question remains whether the normal-state spin fluctuations are simply governed by the Fermi-surface topology, or other effects, such as orbital fluctuations.  Indeed, it has been proposed that orbital-spin fluctuations in a multiband ground state could give rise to the superconducting pairing.\cite{Stanescu:2008p8336,Kontani:2009p8946}  Measurements of magnetic fluctuations in the normal state should provide vital information to resolve these issues.

Hence our goal is to investigate the change in the spin dynamics as a function of doping and elucidate the interconnection with the band structure.  For the electron-doped compounds, recent ARPES and transport studies clearly show the disappearance of the hole pockets around the antiferromagnetic zone center, {\it i. e.},  a Lifshitz transition.\cite{Lifshitz:1960p8962}  This occurs for an electron doping $x_{\rm L}$ of Ba(Fe$_{1-x}$Co$_x$)$_2$As$_2$ ($0.15<x_{\rm L}<0.3$ from ARPES, whereas $x_{\rm L}\sim0.1$  from transport measurements).\cite{Brouet:2009p8677,Sudayama:2010p10433,katayama09} The doping dependence of the spin dynamics, however, has not been comprehensively studied on single-crystal samples by neutron scattering.  Previous work focused on a powder sample\cite{Wakimoto:2009p8895} or on the spin resonance in the optimally doped and underdoped compounds.\cite{Christianson:2008p306, Christianson:2009p7703,Lumsden:2009p543,Li:2009p6138,Qiu:2009p8258,Chi:2009p737}  Here we investigate the spin dynamics in single-crystal, electron-doped Ba(Fe$_{1-x}$Co$_x$)$_2$As$_2$ for $x = 0$, 0.06, 0.14, and 0.24, ranging from the parent compound to the heavily overdoped regime, with a particular emphasis on magnetic fluctuations in the normal state.  We find that the magnetic fluctuations in the paramagnetic state of the parent compound are remarkably similar to those in the normal state of the optimally doped compound.  On the other hand, the spin dynamics in the overdoped regime is drastically different, and resembles that in the overdoped cuprates. As the cobalt content increases well into the heavily overdoped regime, magnetic scattering disappears, coinciding with the absence of superconductivity.

\begin{figure*}
\centering \vspace{0in}
\includegraphics[width=15cm]{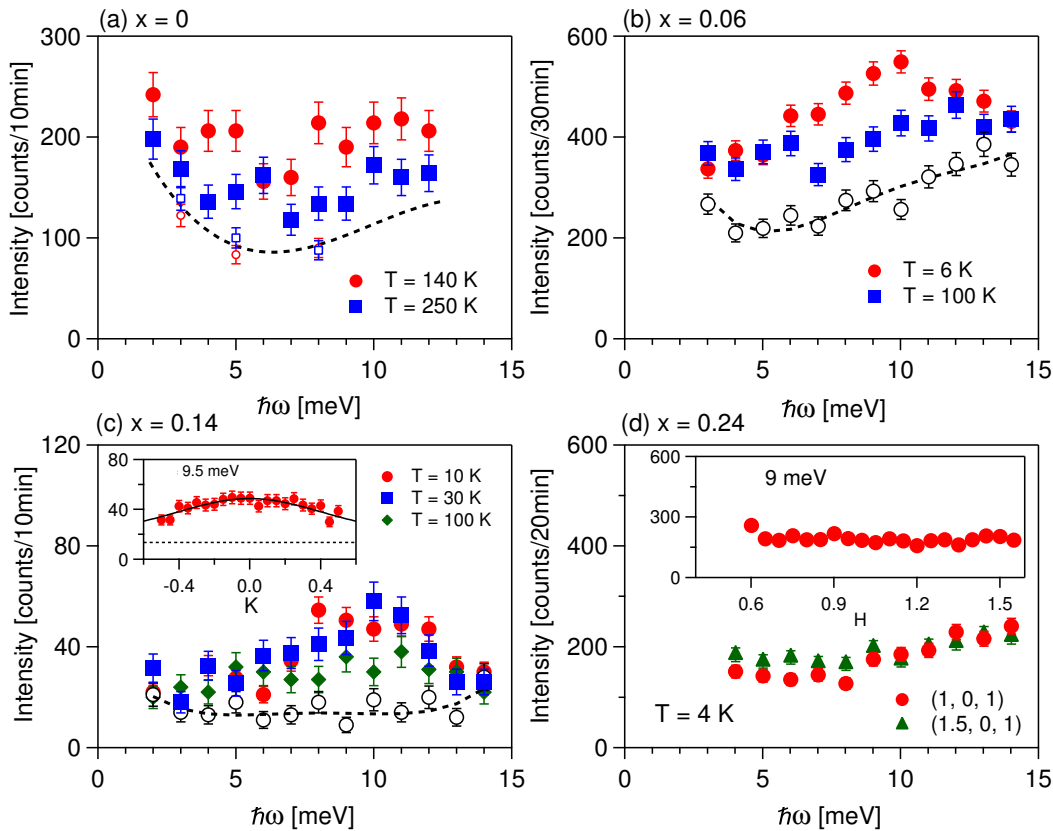}
\caption{(Color online) Constant-Q scans were measured on samples of four cobalt concentrations, (a) $x=0$ at $Q=(1,0,1)$ (BT7), (b) $x=0.06$ at $Q=(1,0,1)$ (HB3), (c) $x=0.14$ at $Q=(1,0,0)$ (GPTAS), and (d) $x=0.24$ (HB3). Background is shown by open symbols.  For $x=0$, background was estimated from constant-energy scans at $\hbar\omega=3$, 5, 8, and 15 meV. In (a), open red circles and open blue squares denote background calculated from constant-energy scans at 140 K and 250 K, respectively. For $x=0.06$ and $0.14$ background was measured away from the peak positions at $Q=(1.2,0,1)$ and $(3,0,0)$, respectively. For $x=0.24$, both constant-Q and constant-energy scans do not show scattering intensity above background, which was measured at $Q=(1.5,0,1)$.  The dotted lines in (a), (b), and (c) denote the fitted background}\label{fig1}
\end{figure*}

\section{Experimental details}
All single-crystal samples were grown from a self-flux using the Bridgman method described in Ref.~\onlinecite{Morinaga:2009p103}. The cobalt content was determined by energy dispersive x-ray analysis (EDX) using a scanning electron microscopy.  Magnetic susceptibility of the $x=0.06$ and $x=0.14$ compounds exhibits superconducting transitions (onset) at 26 K and 7 K, respectively.  The transition temperatures place the former close to the optimally doped regime and the latter in the overdoped regime.  For the $x=0.24$ compound, the superconducting state is not observed down to 1.8 K.  For each composition, up to four single crystals were co-aligned yielding a total mass of about 1 g.  Inelastic neutron scattering measurements were performed on the triple-axis spectrometers BT7 at the NIST Center for Neutron Research, HB3 at the High Flux Isotope Reactor, Oak Ridge National Laboratory, and ISSP-GPTAS at the Japan Atomic Energy Agency.  For clarity, all momentum transfers are labeled using the orthorhombic space group $Fmmm$, a low-temperature phase of the parent compound, even though the proper crystal structure is the tetragonal space group $I4/mmm$. The $x=0$, 0.06, and 0.24 samples were aligned in the $h0l$ zone, while the $x=0.14$ sample was aligned in the $hk0$ zone.  The final neutron energy was fixed at 14.7 meV.  Pyrolytic graphite (PG) crystals were used to monochromate and analyze the incident and scattered beams using the 002 reflection, respectively.  Horizontal collimations of open$-50'-\textrm{sample}-50'-$open, $48'-60'-\textrm{sample}-80'-120'$, and $40'-80'-\textrm{sample}-80'-80'$, were employed at BT7, HB3, and GP-TAS, respectively.  PG filters were placed in the scattered beam.  The samples were cooled using a closed cycle $^4$He cryostat.

\section{Magnetic fluctuations in the normal state}
The scattering intensity can be written as $S_T(Q,\omega)=(n(\omega,T)+1)\chi''_T(\textbf{q},\omega)$, where $n(\omega, T)$ is the Bose factor and $\chi''$ is the imaginary part of the spin susceptibility.  According to the theory of nearly antiferromagnetic metals\cite{moriya}, $\chi''_T(\textbf{q},\omega)$ is given by   
\begin{equation}
\chi''_T(\textbf{q},\omega)=\frac{\chi_0(T)\Gamma(T)\hbar\omega}{(\hbar\omega)^2+\Gamma(T)^2\cdot(1+D^2q_c^2+F^2(q_a^2+q_b^2))^2},\label{eq1}
\end{equation}
where $\Gamma$ is the damping constant, $D$ and $F$ represent the magnetic correlation lengths along the out-of-plane and in-plane directions, respectively, $\chi_0$ represents the isothermal susceptibility, and \textbf{q}=($q_a$,$q_b$,$q_c$) is a wave vector away from an antiferromagnetic zone center.  Background was estimated from a series of constant-energy scans or a constant-$Q$ scan taken away from the antiferromagnetic wave vector and fit to a polynomial function of both momentum transfer and energy transfer.  The coefficients of this polynomial were initially assumed to be temperature dependent.

\begin{figure}
\centering \vspace{0in}
\includegraphics[width=9cm]{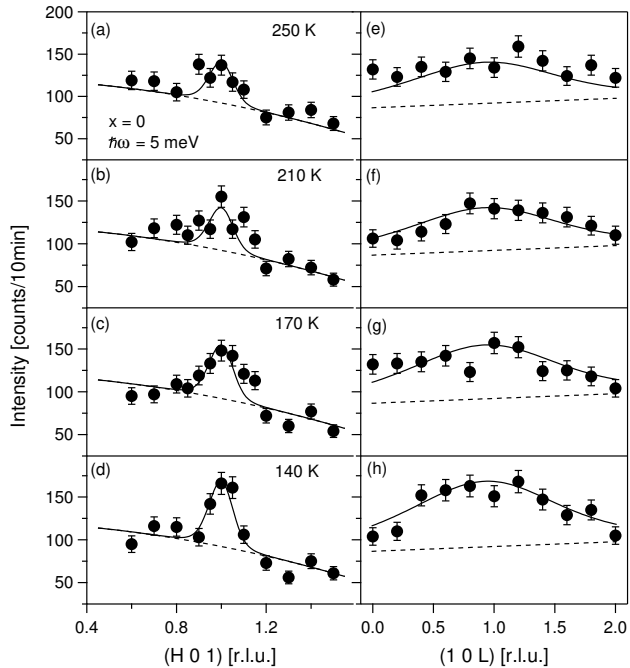}
\caption{Representative in-plane ((a)-(d)) and out-of-plane ((e)-(h)) constant-energy scans at $\hbar\omega=5$ meV in the paramagnetic state of the parent compound. The solid lines denote the results of the global fit.}\label{fig2}
\end{figure}

\begin{figure}
\centering \vspace{0in}
\includegraphics[width=9cm]{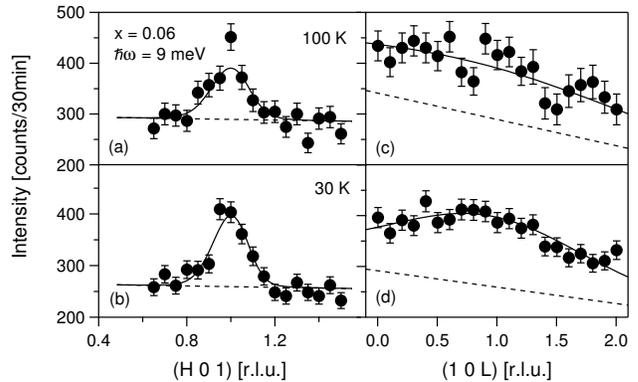}
\caption{Representative in-plane ((a) and (b)) and out-of-plane ((c) and (d)) constant-energy scans at $\hbar\omega=9$ meV in the normal state of the optimally doped compound. The solid lines denote the results of the global fit.}\label{fig3}
\end{figure}

\subsection{Undoped $x=0$}
\begin{figure*}
\centering \vspace{0in}
\includegraphics[width=15cm]{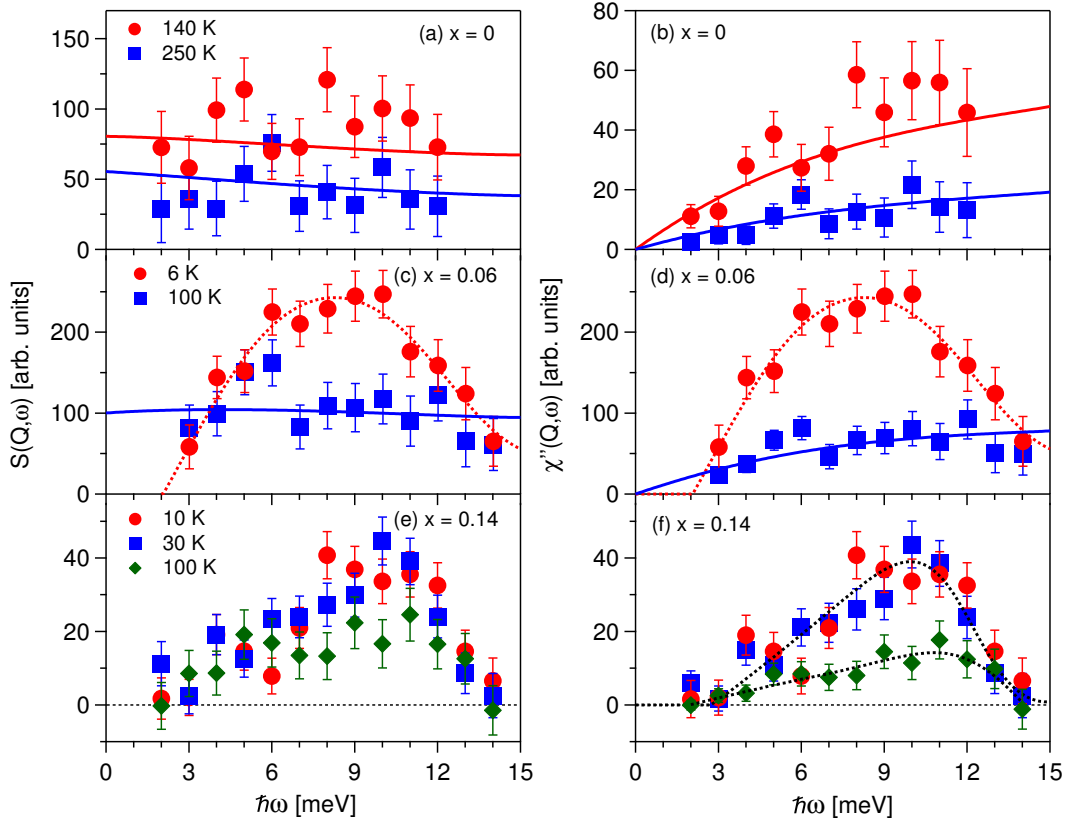}
\caption{(Color online) The background subtracted scattering intensity and the imaginary part of the spin susceptibility $\chi''$ at different cobalt concentrations, (a)-(b) $x=0$ at $Q=(1,0,1)$, (c)-(d) $x=0.06$ at $Q=(1,0,1)$, and (e)-(f) $x=0.14$ at $Q=(1,0,0)$.  The dotted lines are guides to the eye.  The solid lines denote the global fits to Eq.~\ref{eq1} convoluted with the resolution function. We note that $\chi''(Q,\omega)$ is identically zero for $\omega=0$ since it is required to be an odd function of $\omega$.}\label{fig4}
\end{figure*}

\begin{figure}
\centering \vspace{0in}
\includegraphics[width=8.5cm]{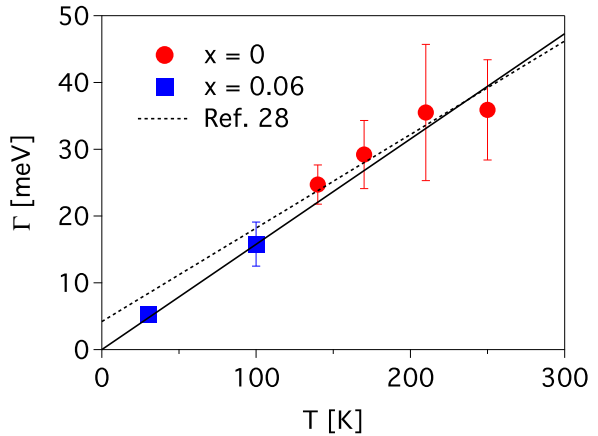}
\caption{(Color online) The damping constant $\Gamma$ as a function of temperature.  Red circles and blue squares represent the damping constants of the parent and optimally doped compounds, respectively.  The dotted line shows the linear relation measured on the optimally doped compound ($x=0.075$)\cite{inosov}.  The solid line denotes our result (see the text).}\label{fig5}
\end{figure}

For the parent compound ($x=0$), our analysis of the background shows weak temperature dependence in the measuring temperature range.  We, therefore, assume temperature-independent background in the fitting (see Figure 2). The background-subtracted scattering intensity $S_T(Q,\omega)$ was then converted to $\chi''$.  Representative constant-Q scans, constant-energy scans, $S_T(Q,\omega)$, and $\chi''$ measured in the paramagnetic state are shown in Figures~\ref{fig1}(a),~\ref{fig2}, and~\ref{fig4}(a)-(b), respectively. $\chi''$ was fitted to the theory of nearly antiferromagnetic metals according to Eq.~(1). In the fitting procedure, $\chi_0$ was constrained to obey the Curie law, since the data at high energy ($\hbar\omega\gtrsim15$ meV), which are required to uniquely determine $\Gamma$ and $\chi_0$ at high temperatures, are not available. Error bars correspond to three times the standard deviation and in Figure~\ref{fig5} indicate large uncertainty at high temperatures, where $\Gamma$ lies beyond the measuring energy range.  We found that $D$ and $F$ do not change significantly within the measuring temperature range between 140 K and 250 K, and the difference lies within the uncertainties. Therefore, we were unable to determine their temperature-dependence and hence their values were fixed at 2.6(5) \AA~and 20(6) \AA, respectively. We note that the uncertainties were obtained from fitting the 140 K data and that they could become larger at high temperatures (see Figure~\ref{fig2}).  We have previously reported these anisotropic magnetic fluctuations in the paramagnetic state of the parent compound at 145 K~\cite{matan}.  We have refitted those data using Eq.~(\ref{eq1}) (not shown), and have found that the lineshapes can be well described by Eq.~(\ref{eq1}) using the same parameters $D$ and $F$, while $\Gamma$ at 145 K was obtained from the linear relation mentioned below (also see Figure~\ref{fig5}). 

At each temperature, the solid lines in Figures~\ref{fig1}(a),~\ref{fig2} and~\ref{fig4}(a)-(b) denote the global fits to several constant-energy scans and the constant-$Q$ scan, convoluted with the four dimensional resolution function.  The resulting fit parameter $\Gamma$ is linearly proportional to temperature, e.g., $\Gamma(T)=\alpha\cdot T$, where $\alpha=0.16(6)$ meV/K as shown by the solid line in Figure~\ref{fig5}.  Note that $\Gamma$ remains finite at the ordering temperature $T_N = 136$ K, and we do not observe any divergence of the correlation length at $T_N$.  Therefore, unlike the parent compounds of the cuprates, the observed magnetically ordered state in BaFe$_2$As$_2$ is most likely not driven by the spin dynamics of the paramagnetic phase, but may be explained in light of the interconnection between the lattice and magnetic interactions.

\subsection{Optimally doped $x=0.06$}
For the optimally-doped compound ($x=0.06$), the antiferromagnetic order is completely suppressed, and superconductivity emerges for $T<T_c=26$ K. In the superconducting state, we observe the broad inelastic scattering centered at $\hbar\omega=9.6$ meV (Figures~\ref{fig1}(b) and~\ref{fig4}(c)-(d)), in agreement with earlier reports,\cite{Lumsden:2009p543,inosov} where the peak is attributed to the resonance mode.  As temperature increases above $T_c$, the inelastic peak is replaced by quasielastic magnetic fluctuations.  Similar to the parent compound, the imaginary part of the spin susceptibility can be well described by the theory of nearly antiferromagnetic metals [Eq. (\ref{eq1})], which is shown by the solid lines in Figures~\ref{fig1}(b),~\ref{fig3}, and~\ref{fig4}(c)-(d). The magnetic correlation length $D$ ($F$) is equal to 2.4(6)~\AA~[19(3)~\AA)] at 30 K and 0.9(9)~\AA~[12(3)~\AA] at 100 K.  If measured at the same temperature, the correlation lengths measured in the optimally doped compound are shorter than those measured in the parent compound, which could be due to the change in the spin concentration upon doping.  The in-plane magnetic correlation length is consistent with the earlier report on the $x=0.075$ compound\cite{inosov}.  We note that the out-of-plane magnetic correlation length was not measured in Ref.~\onlinecite{inosov}. This anisotropic magnetic fluctuations observed in the paramagnetic state of the parent compound and in the normal state of the optimally doped compound are reminiscent of the anisotropic exchange interactions measured in the ordered state of the parent compound.\cite{matan}

We observe a marked similarity in the magnetic fluctuations in the normal state of the optimally doped compound and those in the paramagnetic state of the parent compound. More importantly, the temperature dependence of $\Gamma$ follows the same linear relation as that observed in the parent compound (Figure~\ref{fig5}).  As a comparison, Inosov et al.\cite{inosov} reported that the temperature dependence of $\Gamma$ follows a similar linear form $\Gamma(T)=\alpha\cdot(T+\Theta)$ in the $x=0.075$ compound ($T_c=25$ K), where $\alpha=0.14(4)$ meV/K and $\Theta$, the Curie-Weiss temperature, is equal to 30(10) K. The fact that the imaginary part of the spin susceptibility in the parent and optimally doped compounds can be well described by the theory of nearly antiferromagnetic metals suggest that the magnetic fluctuations in the normal state of the parent and optimally doped compounds could have a common origin, and are likely related to the presence of the quasi-two-dimensional hole and electron pockets observed by photoemission;\cite{Brouet:2009p8677} more evidences of this interconnection will be presented below.

\subsection{Overdoped $x=0.14$}
In contrast to the parent and optimally doped compounds, the spin dynamics in the overdoped, superconducting compounds ($x=0.14$) shows  the depletion of the scattering intensity in the low-energy region and gap-like excitations around 10 meV (Figures~\ref{fig1}(c) and~\ref{fig4}(e)-(f)).  Its peak profile cannot be fit to either Gaussian or Lorentzian line shapes. Furthermore, the scattering intensity exhibits weak temperature dependence, and the in-plane magnetic correlations are much shorter than those of the parent and optimally doped compounds as shown in the inset of Figure~\ref{fig1}(c). The  imaginary part of the spin susceptibility [Figure~\ref{fig4}(f)] displays linear energy-dependence at low energy and a sharp drop at high energy. With the exception of the gap-like behavior, these magnetic excitations share many characteristics with those observed in the overdoped cuprates.\cite{wakimoto}

The origin of the magnetic excitations in the overdoped, superconducting compound ($x=0.14$) is unclear at the present moment.  We note that $x=0.14$ is a higher doping than the neck-collapsing Lifshitz transition ($x\sim0.1$), and is indeed very close to the hole-pocket-vanishing Lifshitz point.\cite{Lifshitz:1960p8962} Thus, it is likely that the dominant part of the hole band lies below the Fermi level.  Recent theoretical calculations show that in such a case the imaginary part of the spin susceptibility is strongly suppressed giving rise to the pseudogap behavior\cite{ikeda09}, and such a pseudogap was observed in NMR measurements in the electron overdoped regimes of LaFeAsO$_{1-x}$F$_x$ (Ref.~\onlinecite{Nakai:2009p8950}) and Ba(Fe$_{1-x}$Co$_x$)$_2$As$_2$.\cite{Ning:2009p8947} Even though the gap-like behavior observed in our neutron scattering suggests that the majority of the hole Fermi surface already diminishes at $x=0.14$, superconductivity with lower $T_{\rm c} = 7$~K is still observed.  This superconductivity may then be of the nodal type, where pairing is formed between electrons on the same electron Fermi surface around the zone corners, as has been theoretically proposed as one alternative to the $s_\pm$ mechanism.\cite{kuroki08}  A recent heat transport experiment indeed indicates that the superconducting gap shows a tendency to be strongly anisotropic as the electron doping increases in Ba(Fe$_{1-x}$Co$_x$)$_2$As$_2$,\cite{tanatar09} although the measurements were made only up to $x=0.114$ so a direct comparison with our result ($x = 0.14$) is not possible at present. 

\subsection{Heavily overdoped $x=0.24$}
For the heavily overdoped, non-superconducting compound ($x=0.24$), Figure~\ref{fig1}(d) shows the suppression of the magnetic scattering. NMR measurements on the $x=0.26$ compound\cite{Ning:2009p8947} and neutron scattering measurements on electron-doped LaFeAsO$_{1-x}$O$_x$ in the heavily overdoped regime\cite{Wakimoto:2009p8895} reveal the suppression of the spin fluctuations consistent with our result.  In this regime, both photoemission measurements and first-principles calculations point to the disappearance of a hole Fermi-surface pocket.\cite{Brouet:2009p8677,Stanescu:2008p8336} Our result, therefore, further suggests the correlation between the electronic band structure and magnetism, and supports the scenario that the magnetic fluctuations in the underdoped and optimally doped regimes, which serve as a precursor to superconductivity, originate from quasiparticle scattering across the electron and hole pockets.\cite{Mazin:2009p6797}

\section{Conclusion}
We have studied the spin dynamics in electron-doped Ba(Fe$_{1-x}$Co$_x$)$_2$As$_2$ at four cobalt concentrations. We observe a striking similarity between the magnetic fluctuations in the paramagnetic state of the parent compound and those in the normal state of the optimally doped compound, and the suppression of the magnetic signal in the heavily overdoped regime, in which superconductivity disappears.  These two results suggest that magnetism and superconductivity are strongly correlated.  On the other hand, magnetic excitations in the vicinity of the hole-pocket-vanishing Lifshitz point are markedly different from those in the underdoped regime, with the emergence of a spin gap and much weaker temperature dependence of $\chi''$.  These changes in the spin dynamics at different doping levels likely reflect changes in the electronic band structure.  Further experimental and theoretical work is desirable to examine the interconnection between the band structure, spin fluctuations, and superconducting pairing mechanism.

\begin{acknowledgments}
We thank H. Yoshizawa, T. Mizokawa, H. Ikeda, K. Ohgushi, and K. Ishida for valuable discussions.  This work is partly supported by the U.S.-Japan cooperative program on neutron scattering research. Part of this work was supported by the Division of Scientific User Facilities, Office of Basic Energy Sciences, U.S. DOE.
\end{acknowledgments}

\bibliography{BaFe2As2-spinDynamics}

\end{document}